\documentclass[twocolumn]{svjour3}
\pdfoutput=1

\usepackage{url}
\usepackage{color}
\usepackage[pdftex]{graphicx}
  \usepackage{amsmath}
\usepackage{amsfonts}
\makeatletter \let\cl@chapter\relax \makeatother

 \usepackage{hyperref}

\newcommand{\lb}{\left[}
\newcommand{\rb}{\right]}

\newcommand{\be}{\begin{equation}}
\newcommand{\ee}{\end{equation}}

\newcommand{\al}{\alpha}

\newcommand{\sdg}{\sqrt{-g}}
\newcommand{\sdf}{\sqrt{-f}}
\newcommand{\sdh}{\sqrt{-h}}
\newcommand{\mn}{{\mu\nu}}
\newcommand{\ab}{{\alpha\beta}}

\newcommand{\ag}{\alpha_g}
\newcommand{\af}{\alpha_f}

\begin{document}

\title{The nature of spacetime in bigravity}
\subtitle{Two metrics or none?}

\author{Yashar Akrami \and Tomi S. Koivisto \and Adam R. Solomon}

\institute{Y. Akrami \at Institute of Theoretical Astrophysics, University of Oslo\\
       P.O. Box 1029 Blindern, N-0315 Oslo, Norway \\\email{yashar.akrami@astro.uio.no} \and T. S. Koivisto \at Nordita, KTH Royal Institute of Technology and Stockholm University\\
       Roslagstullsbacken 23, SE-10691 Stockholm, Sweden\\ \email{tomi.koivisto@fys.uio.no} \and A. R. Solomon \at DAMTP, Centre for Mathematical Sciences, University of Cambridge\\
       Wilberforce Rd., Cambridge CB3 0WA, UK \\\email{a.r.solomon@damtp.cam.ac.uk}}

\date{\today}

\maketitle

\begin{abstract}
The possibility of matter coupling to two metrics at once is considered. This appears natural in the most general ghost-free, bimetric theory of gravity, where it unlocks an additional symmetry with respect to the exchange of the metrics. This double coupling, however, raises the problem of identifying the observables of the theory. It is shown that if the two metrics couple minimally to matter, then there is no physical metric to which all matter would universally couple, and that moreover such an effective metric generically does not exist even for an individual matter species. By studying point particle dynamics, a resolution is suggested in the context of Finsler geometry.
\end{abstract}


The seemingly innocuous modification of Einstein's general theory of relativity (GR) by adding a tiny mass to the graviton may entail radically new perspectives on the nature of spacetime and its interplay with matter. In the context of GR, the graviton is identified with the fluctuations of the metric describing the curvature of spacetime. To give these fluctuations a mass, we need to introduce an external metric and are thus led to consider \textit{bigravity}, where matter can couple to two distinct metrics. Consequently, the traditional notion of a ``physical metric'' may have to be discarded, leaving us faced with entirely new conceptual challenges in interpreting even the observables of the theory. 

This is what we pursue in this report, by studying a version of bimetric massive gravity where both metrics are minimally coupled to matter. During the past few years, following the discovery of the ghost-free, nonlinear completion of the Fierz-Pauli action for linearized massive gravity \cite{Fierz_et_Pauli1939} by de Rham, Gabadaze and Tolley \cite{deRham_et_Gabadadze2010b,deRham_Gabadadze_et_Tolley2010} and its promotion to a bimetric theory by Hassan and Rosen \cite{Hassan_et_Rosen2011a,Hassan:2011hr,Hassan_et_al2012a,Hassan_et_Rosen2012a,Hassan:2011ea}, the pile of papers investigating bimetric theories has swelled (see, e.g., \cite{deRham:2014zqa} for a review), but the possibility of matter becoming ``doubly coupled'' has been neglected in the literature apart from a handful of exceptions \cite{Hassan_et_Rosen2012a,Khosravi:2011zi,Hassan:2012wr,Akrami:2013ffa,Tamanini:2013xia}. If matter couples only to one of the metrics, then that retains its usual role as the physical metric, while the other spin-2 particle can be interpreted as extra matter whose interactions with the graviton give it mass.

It is therefore helpful to begin by clearly spelling out our motivations for inquiring into the unconventional ``doubly-coupled'' spacetimes. Firstly, the structure of the viable bigravity action is symmetric under exchange of the two metrics. We also note that one could arrive at the double minimal coupling from extra-dimensional constructions \cite{deRham:2013awa} attempting to obtain bigravity as the low-energy limit. However, it was very recently shown \cite{Yamashita:2014fga,deRham:2014naa} that minimally coupling matter to each metric separately, which is the case we consider here, revives the Boulware-Deser ghost \cite{Boulware:1973my}. Therefore the following action cannot be considered a suggestion for a fundamental theory of bimetric massive gravity. Nonetheless, it provides a simple and illustrative example of the dynamics in a theory which lacks an effective metric for matter, and should be considered a proof of concept. Note that there seem to be alternative ways of coupling the two metrics to matter simultaneously without reintroducing the ghost, as is suggested by examples in Ref. \cite{deRham:2014naa}. In those cases an effective metric may exist, in which case the issues raised in this report would find obvious resolutions. It is however not clear yet whether such theories are immune from other types of pathologies. It can very well be the case that the heretofore-unknown, healthy doubly-coupled theory of gravity will not admit an effective-metric formulation. It is the aim of the present investigation to demonstrate the difficulties such theories would have with regards to defining observables, by studying arguably the simplest example of doubly-coupled bigravity without an effective-metric description.

In the theory we consider, the full action for the two metrics ($g_{\mu\nu}$ and $f_{\mu\nu}$) and the matter fields (denoted collectively by $\Psi$) is
\begin{align} \label{action}
  S &= \int d^4x\sqrt{-g}\lb \frac{M^2_g}{2}R(g)  - \al_g \mathcal{L}_m\left(g, \Psi\right)\rb \nonumber \\
            &\hphantom{{}=}- \int d^4x\sqrt{- f}\lb \frac{M^2_f}{2}R(f)   - \af \mathcal{L}_m\left(f, \Psi\right)\rb \nonumber \\ 
 &\hphantom{{}=}+  m^2M_g^2\int d^4x\sqrt{-g}\sum_{n=0}^{4}\beta_ne_n\left(\sqrt{g^{\mu\alpha}f_{\alpha\nu}}\right)\,.
\end{align}
The first two lines include Einstein-Hilbert terms for each metric (with coupling constants $M_g$ and $M_f$, respectively) and minimally-coupled matter (with coupling strengths $\al_g$ and $\al_f$, respectively). The third line is the potential, comprising the five possible ghost-free interaction terms for the two metrics \cite{deRham_Gabadadze_et_Tolley2010,Hassan_et_Rosen2011a,Hassan_et_Rosen2012a}. Due to the double coupling, this action (including the potential) is symmetric under $g_\mn \Leftrightarrow f_\mn$ with the parameter exchanges
\begin{equation}
M_g \Leftrightarrow M_f\,, \qquad
\al_g \Leftrightarrow \al_f\,, \qquad
\beta_i \Leftrightarrow \beta_{4-i}\,. 
\end{equation}
By studying the Bianchi identities we see that the presence of interactions between the metrics is crucial for the existence of nontrivial solutions; it is not viable to couple two pure, non-interacting GR sectors to matter \cite{Akrami:2013ffa}.

An immediate concern is the violation of the equivalence principle. However, because the Vainshtein mechanism screens massive gravity effects \cite{Vainshtein:1972sx}, it is not obvious how stringent constraints we could obtain from tests of GR in the solar system: the modifications can be hidden from local experiments while showing up at cosmological scales. The cosmology of this doubly-coupled theory has been studied and shown to produce viable late-time accelerating background expansion without an explicit cosmological constant term \cite{Akrami:2013ffa}, and with a phenomenology which can be interestingly different from that of the singly-coupled theory \cite{Akrami:2012vf}. The implications for large-scale structure formation \cite{Solomon:2014dua,Konnig:2014xva} have not been detailed in the presence of such a double coupling.

In this report we focus on the behavior of matter minimally coupled to two metrics. At the level of cosmological background expansion, it turns out that dust-like matter obeys the conservation equations for both of the metrics simultaneously; moreover, for several solutions of interest, the two independent metrics become conformally related, rendering the identification of the observables in the theory more straightforward \cite{Akrami:2013ffa}. In general this is not the case. In fact, there is generically no effective physical metric in terms of which to interpret the theory by direct analogy to GR. A novel insight we will arrive at here is that despite the purely Riemannian starting point (\ref{action}), matter effectively lives in a Finslerian spacetime.

We can readily confirm that no physical Riemannian metric exists in the sense that matter would minimally couple to it and thus follow its geodesics. Consider the electromagnetic field $A_\mu$. This is of paramount importance for cosmology, since we make observations by tracking photons. Its action is
\begin{align}
S_A &= \frac{1}{4}\ag \int d^4x \sdg g^{\mu\alpha}g^{\nu\beta}F_\mn F_\ab\nonumber \\
&\hphantom{{}=}- \frac{1}{4}\af \int d^4x \sdf f^{\mu\alpha}f^{\nu\beta}F_\mn F_\ab, \label{eq:maxwellaction}
\end{align}
where $F_\mn = \partial_\mu A_\nu - \partial_\nu A_\mu$ does not depend on a metric. We consider $A_\mu$ to be minimally coupled to an effective metric $h_{\mu\nu}(g,f)$ if Eq.~(\ref{eq:maxwellaction}) can be written
\begin{equation}
S_A = -\frac{1}{4} \int d^4x \sdh h^{\mu\alpha}h^{\nu\beta}F_\mn F_\ab.
\end{equation}
This implies that $h_\mn$ obeys
\begin{equation}
\ag \sdg g^\mn g^\ab + \af \sdf f^\mn f^\ab = \sdh h^\mn h^\ab.
\end{equation}
However, this equation overconstrains $h_\mn$: in general, it cannot simultaneously satisfy the $00-00$, $00-ii$, and $ii-ii$ components. Thus there is no physical metric for the electromagnetic field.


Similar arguments hold for other fields, such as a massive scalar. A massless scalar does have an effective metric, defined by
\begin{equation}
\sdh h^\mn = \ag \sdg g^\mn + \af \sdf f^\mn.
\end{equation}
But a scalar with a non-trivial potential cannot, in general, minimally couple to a metric $h_\mn$ without inducing either a non-canonical kinetic term or a spacetime-varying mass in the effective Lagrangian $\mathcal{L}_m\left(h, \phi\right)$. For some special choices of $g_\mn$ and $f_\mn$, particularly if they are related by a constant conformal factor, the mass or kinetic term would only rescale by a constant amount, but such a relation between $g_\mn$ and $f_\mn$ is far from general.

This situation is radically different from the extensively-studied nonminimally coupled theories where the behavior of matter can be described in terms of a single metric. In the context of scalar-tensor theories, for example, it is well known that there are conformally-equivalent descriptions of the theory where either the gravity sector is GR whilst matter has a nonminimal coupling (the Einstein frame), or matter is minimally coupled whilst the gravity sector is modified (the Jordan frame). All physical predictions are completely independent of the frame in which they are calculated after properly taking into account the rescaling of units in the Einstein frame, as explained with depth and clarity in the seminal paper of Brans and Dicke \cite{Brans:1961sx}. One can generalize to non-universal couplings, allowing different Jordan frame metrics for different matter species, or to couplings to multiple fields. These bring about new technical but not fundamental difficulties. However, the bigravity theories (\ref{action}) do not admit a Jordan frame at all for most types of matter. They possess mathematically two metrics but physically none, and to understand them we need to step beyond the confines of metric geometry. 

For concreteness, let us look at the simplest possible type of matter: a point particle of mass $m$. Its action is defined by
\begin{align}
S_\mathrm{pp} &= -m\ag \int dt \sqrt{g_\mn \dot x^\mu \dot x^\nu} - m \af \int dt \sqrt{f_\mn \dot x^\mu \dot x^\nu} \nonumber \\
&= -m \ag \int ds_g - m \ag \int ds_f,
\end{align}
where $ds_g^2 = g_\mn dx^\mu dx^\nu$, $ds_f ^2= f_\mn dx^\mu dx^\nu$, and overdots denote derivatives with respect to a parameter $\lambda$ along the particle's trajectory, $x^\mu(\lambda)$. The ``geodesic" equation we derive from this is
\begin{align}
&\ag g_\ab\left(\frac{du_g^\alpha}{ds_g} +\overset{g}{\Gamma}\vphantom{\Gamma}^\alpha_{\mu\nu}u_g^\mu u_g^\nu\right) \nonumber \\
+&{} \af f_\ab\frac{ds_f}{ds_g}\left(\frac{du_f^\alpha}{ds_f} +\overset{f}{\Gamma}\vphantom{\Gamma}^\alpha_{\mu\nu}u_f^\mu u_f^\nu\right) = 0\,,  \label{eq:ppeom}
\end{align}
where $u_g^\mu \equiv dx^\mu/ds_g$ is the four-velocity properly normalized with respect to $g_\mn$, such that $g_{\mu\nu}u_g^\mu u_g^\nu=1$, and $u_f^\mu$ is defined analogously for the $f_\mn$ geometry.

Equation (\ref{eq:ppeom}) is not the geodesic equation for any Riemannian metric. To see this, let us assume there is a line element $ds$ for which $S_\mathrm{pp} = -m\int ds$. This implies
\begin{align}
ds^2 & = \left(\ag^2g_\mn + \af^2f_\mn\right)dx^\mu dx^\nu \nonumber \\
&\hphantom{{}=}+ 2\alpha_g\alpha_f\sqrt{g_\mn f_\ab dx^\mu dx^\nu dx^\alpha dx^\beta}. \label{eq:ppline}
\end{align}
Eq.~(\ref{eq:ppline}) is in fact the line element of a \textit{Finsler geometry} \cite{Cartan:1934,Bekenstein:1992pj}, the most general line element that is homogeneous of degree 2 in the coordinate intervals $dx^\mu$,
\begin{equation}
ds^2 = f(x^\mu,dx^\nu);\qquad f(x^\mu,\lambda dx^\nu) = \lambda^2f(x^\mu,dx^\nu). \label{eq:finslerdef}
\end{equation}
Using Euler's theorem for homogeneous functions, the line element can be defined in terms of a \textit{quasimetric} $\mathcal{G}_\mn$,
\begin{equation}
f = ds^2 = \mathcal{G}_{\mu\nu}dx^\mu dx^\nu\,, \quad
\mathcal{G}_{\mu\nu} = \frac{1}{2}\frac{\partial^2 f}{\partial dx^\mu \partial dx^\nu}\,. \label{eq:qm}
\end{equation}
Note, however, that the quasimetric $\mathcal{G}_{\mu\nu}$ can depend on $dx^\mu$, which is how it differs from the metric of a usual Riemannian spacetime. We find that $\mathcal{G}_{\mu\nu}$ can be written as
\begin{align}
\mathcal{G}_{\mu\nu}  &=  \ag^2g_\ab \af^2f_\ab  + \ag\af\Big[\frac{ds_f}{ds_g}\left(g_{\mu\nu} - u^g_\mu u^g_\nu\right) \nonumber \\
&\hphantom{{}=}+ \frac{ds_g}{ds_f}\left(f_\ab - u^f_\mu u^f_\nu \right) + 2u^g_{(\mu}u^f_{\nu)}\Big]\,.
\end{align}
This quasimetric is disformally related \cite{Bekenstein:1992pj} to the original metrics.

Defining the proper time $\tau$ by $d\tau^2 = -ds^2$, it is easy to see that massive point particles travel on unit-norm timelike geodesics with respect to the quasimetric $\mathcal{G}_{\mu\nu}$. By using the same einbein trick with a Lagrange multiplier as in GR, we can now also extend this treatment to massless particles, and find that they travel along null geodesics of the same quasimetric $\mathcal{G}_{\mu\nu}$. We note in passing that these results can be straightforwardly generalized to theories with more than two metrics.

The investigation of spacetimes endowed with two metrics is rendered quite topical by the recent discovery of the ghost-free family of bigravity actions \cite{Hassan_et_Rosen2012a}.
Here we have considered the fundamental issue of observables in a theory where matter couples to two metrics simultaneously. We focused on a particularly simple example of doubly-coupled theories, in which matter minimally couples to both metrics. We arrived at the result that the geometry that emerges for an observer in such a bimetric spacetime depends quite nontrivially upon, in addition to the two metric structures, the observer's four-velocity. This means she is disformally coupled to her own four-velocity, and thus effectively lives in a Finslerian spacetime. Paradoxically, once we have doubled the geometry, we lose the ability to use its familiar methods. Unless a heathy theory of doubly-coupled bimetric gravity with an effective-metric description exists, this is a call to go back to the basics, and rediscover the justifications for results which we have taken for granted over the better part of the last century.

\begin{acknowledgements}

We thank John Barrow and Fawad Hassan for useful discussions. This work has benefitted from Y. Akrami's and A. Solomon's stays at the University of Oslo and Nordita, as well as at the Structure of Gravity and Space-Time conference in Oxford and from many enlightening discussions therein. Y. Akrami is supported by the European Research Council (ERC) Starting Grant StG2010-257080. A. Solomon acknowledges support from the David Gledhill Research Studentship, Sidney Sussex College, University of Cambridge; and the Isaac Newton Fund and Studentships, University of Cambridge. 

\end{acknowledgements}

\bibliography{essayrefs}

\end{document}